**CancerBERT: A Cancer Domain Specific Language Model for Extracting Breast Cancer Phenotypes from Electronic Health Records**

Sicheng Zhou, MS, Institute for Health Informatics, University of Minnesota, Minneapolis, MN, USA

Nan Wang, School of Statistics, University of Minnesota, Minneapolis, MN, USA

Liwei Wang, MD, PhD, Department of AI and Informatics Research, Mayo Clinic, Rochester, MN, USA

Hongfang Liu, PhD, Department of AI and Informatics Research, Mayo Clinic, Rochester, MN, USA,

Rui Zhang, PhD, Institute for Health Informatics, and Department of Pharmaceutical Care & Health Systems, University of Minnesota, Minneapolis, MN, USA

Corresponding author & address: Rui Zhang, 8-100 Phillips-Wangensteen Building, 516 Delaware Street SE, Minneapolis, MN 55455
Contact information:
Email: zhan1386@umn.edu
Phone number: 6126264209

**Abstract**

**Objective**: Accurate extraction of breast cancer patients' phenotypes is important for clinical decision support and clinical research. This study developed and evaluated cancer domain pre-trained CancerBERT models for extracting breast cancer phenotypes from clinical texts. We also investigated the effect of customized cancer-related vocabulary on the performance of CancerBERT models.

**Materials and Methods**: A cancer-related corpus from breast cancer patients was extracted from the electronic health records of a local hospital. We annotated named entities in 200 pathology reports and 50 clinical notes for eight cancer phenotypes for fine-tuning and evaluation. We kept pre-training the BlueBERT model on the cancer corpus with expanded vocabularies (using both term frequency-based and manually reviewed methods) to obtain CancerBERT models. The CancerBERT models were evaluated and compared with other baseline models on the cancer phenotype extraction task.

**Results**: All CancerBERT models outperformed all other models on the cancer phenotyping NER task. Both CancerBERT models with customized vocabularies outperformed the CancerBERT with the original BERT vocabulary. The CancerBERT model with manually reviewed customized vocabulary achieved the best performance with macro-F1 scores equal to 0.876 (95% CI, 0.873-0.879) and 0.904 (95% CI, 0.902-0.906) for exact match and lenient match, respectively.

**Conclusions**: The CancerBERT models were developed to extract the cancer phenotypes in clinical notes and pathology reports. The results validated that using customized vocabulary may further improve the performances of domain specific BERT models in clinical NLP tasks. The CancerBERT models developed in the study would further help clinical decision support.

**INTRODUCTION**

Breast cancer is one of the most prevalent and lethal cancers for women in the US. It is estimated that there will be about 250,000 patients diagnosed with breast cancer each year and around 40,000 deaths due to breast cancer [1]. The development of precision medicine has contributed new approaches to the better diagnosis, prognosis and treatments of breast cancers, with the ultimate goal of selecting optimal treatments for individual patients [2-4]. A representative example is the targeted therapy for breast cancer, which uses different medications to treat patients with different hormone receptors status, such as human epidermal growth factor receptor 2 (HER2) and estrogen receptor (ER). The application of precision medicine and its related translational research need the support of large amounts of cancer-specific patient clinical information. The widely adopted electronic health record systems (EHRs) are fundamental sources to provide longitudinal and multi-perspective patient clinical data, which includes patients' demographics, lab results, disease progress, treatments, and outcomes. Some of these data are stored in the codified (structured) part of EHRs; however, a large amount of information is distributed in the narrative text data, such as the clinical notes and lab reports [5]. How to effectively extract the target information from the narrative text part of the EHRs data remains an important research topic [6-7]. Previous studies (detailed in Background section) focused mainly on the rule-based and conventional machine learning methods. The state-of-the-art language models such as bidirectional encoder representations from transformers (BERT) [8] have shown significant improvement in many NLP tasks; however, there is no cancer domain specific BERT model for downstream clinical NLP tasks, such as cancer phenotype extraction. In addition, there is no investigation of out-of-vocabulary (OOV) issue for BERT-based model in cancer domain.

**Objective**

To address these gaps in current status of the cancer phenotyping extraction, our contributions in this study include:

1) We developed and evaluated cancer domain specific BERT models (CancerBERT) that are able to extract comprehensive collections of breast cancer related phenotypes (i.e., *Hormone receptor type, Hormone receptor status, Tumor size, Tumor site, Cancer grade, Histological type, Cancer laterality* and *Cancer stage*) from both clinical notes and pathology reports in EHRs. Our CancerBERT models significantly outperformed other existing BERT-based models (e.g., BlueBERT, BioBERT, CharBERT) on our name entity recognition (NER) task to extract target cancer phenotypes for breast cancer patients.

2) We also evaluated different methods to address the OOV issue of original BERT models. Specifically, we used two approaches (domain knowledge-based and statistics-based) to generate and add additional cancer specific words that were missing in the original BERT vocabulary. We found that additional cancer specific words can further improve performance of the CancerBERT model on the NER task.

**Background**

Previous works have developed different approaches to extract information from narrative data in EHRs. Manual chart review is a feasible approach to extract the phenotypes from the clinical texts; however, it is time-consuming and not cost-effective [9-10]. Researchers have developed approaches based on natural language processing (NLP) to finish the task automatically. Before the deep learning era, established studies mainly focused on rule-based, traditional machine learning-based methods, depending on the characteristic of the data and specific tasks. Nguyen, et al. developed a rule-based pipeline in 2015 to extract cancer-related phenotypes, including histological type, cancer grade, primary site and laterality, from the textual contents of the pathology reports in EHRs [11]. The F1 scores for different variables range from 0.61 to 0.93, and a message producer/consumer module was integrated into the pipeline to enable the real-time processing of the reports. Yala, et al. developed a machine learning algorithm to classify phenotypes of breast cancers. N-grams were used as features and boosting algorithm was applied to do the classification of phenotype status. The performances were robust, with F1 scores ranging from 0.57 to 1 for different categories. However, the pipeline was designed for judging the status of phenotypes, mainly binary classification; no detail information in the text can be captured [12]. The DeepPhe software was developed in 2017 to extract cancer phenotypes from clinical records [13]. It could extract a wide range of breast cancer phenotypes from the EHRs through different approaches, such as, rules, domain knowledge bases, and machine learning methods. The inter-annotator agreement (IAA) of the DeepPhe range from 0.2 to 0.96 [13]. Qiu, et al. developed a convolutional neural network (CNN) model to extract the cancer primary site from pathology reports, and the CNN model outperformed the traditional frequency vector space approach with a micro-F score of 0.722 [14]. A coarse-to-fine multi-task CNN model was further proposed to extract the cancer primary site, laterality and grade from the pathology reports at the same time; this model obtained an F-1 score of 0.775 for extracting cancer primary site [15].

These studies focused mainly on the rule-based and conventional machine learning methods. The latest BERT-based models have been developed in recent years and show great advantages in NLP tasks compared to the traditional feature-based machine learning approaches [8]. For BERT-based models, previous works have shown that using biomedical domain specific text as training data can obtain better performance compared to models trained on general-domain language for tasks related to the biomedical domain [16-19]. In clinical domain, studies have explored using the advanced BERT-based models to solve clinical information extraction tasks [16-17,20]. These studies are mainly focusing on testing the BERT model on clinical benchmark datasets, such as Informatics for Integrating Biology and the Bedside (i2b2), SemEval, MedSTS. Currently, only few studies have applied the advanced deep learning models include BERT to extract the cancer phenotypes for cancer patients. For example, BERT models were applied to extract the clinical information for breast cancer patients from Chinese clinical texts and achieved F1 scores of 0.786 to 1 for different clinical concepts [21]. It is known that Chinese and English are different, and there are no existing studies that have explored the BERT-based models to extract cancer phenotypes from clinical notes and pathology reports in English. Furthermore, most BERT-based models deal with the out-of-vocabulary (OOV) issue by tokenizing an unknown word into multiple sub-words that exist in the vocabulary. In this case, the sub-word representations may not capture the semantics of the whole word [22]. Several studies explored to improve the OOV issue by either using character-level word embedding [22-23] or building a brand-new domain specific vocabulary to best match the training corpus [24]. These studies obtained promising results in benchmark tasks, but they need to train the models from scratch, which need a huge training corpus and computing sources.

**METHODS**

**Data collection and annotation**

The data used in this study was obtained from the EHRs of the University of Minnesota (UMN) Clinical Data Repository (CDR). The EHRs of UMN contain the health records of 21,291 breast cancer patients from year 2001 to 2018. We obtained the data with the approval of UMN Institutional Review Board (IRB) under #1210M22601.

To develop the phenotype extraction algorithms, the reference standards (i.e., an annotated corpus) needed to be obtained through chart review. To obtain the standard annotations, an annotation guideline was first developed through iterative discussions. The UMN team reviewed some of the pathology reports and clinical notes to collect the different descriptions of the phenotypes in the EHRs and formed the annotation guideline. We randomly sampled 200 pathology reports and 50 clinical notes of breast cancer patients that contain 9,685 sentences; 221,356 tokens were manually annotated by two annotators (graduate students with clinical or pharmacy background). The target entities were annotated in entity level. Cohen's kappa scores were calculated to ensure the consistency between the annotators. INCEpTION was used as the annotation tool [25]. **Figure 1** shows several examples of the annotation.

Figure 1. Examples of annotation in INCEpION.

We focused on eight breast cancer phenotypes that describe the characteristics of breast cancer, including the *Hormone receptor type, Hormone receptor status, Tumor size, Tumor site, Cancer grade, Histological type, Cancer laterality* and *Cancer stage*. Targeted breast cancer phenotypes, their potential values and the according examples in clinical text are shown in **Table 1**.

Table 1– Breast cancer phenotypes, their potential values and examples in clinical texts.

| Phenotypes | Values | Examples of descriptions in clinical text |
|---|---|---|
| Hormone receptor type | positive, negative | HER2 gene was amplified; Estrogen receptor: Positive (95%, strong staining); Tumor is PR negative (0% staining) |
| Tumor size | numeric values describe volumes | Tumor size: 1.0 x 0.5 x 0.7 cm |
| Tumor site | description of positions | Tumor is at 12 o'clock position and 2 cm from the nipple. |
| Cancer grade | numerical values: (1-3) | Histologic grade: 1 of 3; Sample shows Nottingham grade 2 lesions |
| Histological type | ductal carcinoma in situ (DCIS); lobular carcinoma in situ (LCIS), etc. | Histologic Type of Invasive Carcinoma: ductal carcinoma in situ |
| Cancer laterality | right, left | Specimen Laterality: Right breast; Laterality: - Left TUMOR |
| Cancer stage | TNM staging: TX, Tis, T1-4; NX, N0, N1-3; M0, M1 | pathologic stage is pT4 NX MX |

In total, 200 pathology reports and 50 clinical notes were annotated by two annotators, the Cohen kappa score for annotations was calculated to be 0.91. The annotation statistics are shown in **Table 2**.

*Table 2– Annotation statistics*

|  |  | Total Number | Total Unique Entities |
|---|---|---|---|
| Annotated statistics | Documents | 200 | NA |
|  | Total sentences | 9685 | NA |
|  | Total tokens | 221356 | NA |
| Name entity statistics | Hormone receptor type | 1673 | 29 |
|  | Hormone receptor status | 436 | 14 |
|  | Tumor size | 540 | 305 |
|  | Tumor site | 329 | 173 |
|  | Cancer grade | 271 | 15 |
|  | Cancer laterality | 1192 | 4 |
|  | Cancer stage | 173 | 38 |
|  | Histological type | 1070 | 95 |

**Our model: CancerBERT**

*Pre-training on cancer domain-specific corpus*

In this study, we trained cancer domain-specific BERT models (CancerBERT) that are expected to better capture the semantics in cancer specific-clinical notes and pathology reports, thus improving the performance of the task for extracting breast cancer-related phenotypes. The CancerBERT models training process is illustrated in **Figure 2**.

Figure 2. The training process of CancerBERT models. The CancerBERT models were pre-trained based on the BlueBERT model. The process in the red box was implemented in this study.

The BERT-origin model was trained using Wikipedia and book corpus [15] using the original BERT vocabulary, which was also generated from Wikipedia and book corpus. The BlueBERT was further pre-trained on PubMed and MIMIC III data based on the BERT-origin model using the same vocabulary [17]. We kept pre-training the CancerBERT models based on the BlueBERT model. For the pre-training corpus, we extracted the 4,543,184 clinical notes and 1,278,805 pathology reports (about 1 billion tokens) for 21,291 breast cancer patients from UMN EHRs. The corpus was changed to lower case, no other pre-processing was needed. Hereinafter, we called the CancerBERT trained with original BERT vocabulary as CancerBERT$_{OrigVoc}$ to differentiate other CancerBERT variants using customized vocabulary described below.

*Constructing cancer domain specific vocabulary for improving CancerBERT$_{OrigVoc}$ models*

As outlined above, the vocabulary of CancerBERT$_{OrigVoc}$ is identical to the BERT-origin model [15]; thus, many special words and abbreviations in the clinical narratives cannot be covered. The OOV issue may influence the performance of the language model. In the BERT-based models, the WordPiece tokenizer [26] was applied to deal with the OOV issue. It tokenizes an unknown word into multiple sub-words that exist in the vocabulary. For instance, the word "HER2", a breast cancer-related cell receptor gene, is not in the original BERT vocabulary. It will be tokenized into "HER" and "2" by the WordPiece tokenizer; and the model will then use the average (or the first part) of their word embeddings to represent "HER2". However, the word embeddings of "HER" and "2" cannot correctly represent the semantics of the term "HER2". Thus, it may be helpful to add the unknown word "HER2" into the original vocabulary to train its own word embeddings. Thus, we further explored different approaches to generate and incorporate cancer-specific vocabulary in an attempt to further improve the performance of the CancerBERT$_{OrigVoc}$ model.

In this study, we explored two ways to generate additional cancer-related vocabularies to include in the original BERT vocabulary.

Method 1 - Domain knowledge-based:

    1) SpaCy [27] tokenizer was used to tokenize the breast cancer training corpus from EHRs to produce a new list of words

    2) All unique words in the new word list that did not appear in the original vocabulary were identified

    3) A researcher with clinical background reviewed the newly identified words and selected 397 cancer-related words for vocabulary expansion.

Method 2 - Frequency-based:

    1) SpaCy [27] tokenizer was used to tokenize the breast cancer training corpus from EHRs to produce a new list of words

    2) All unique words in the new word list that did not appear in the original vocabulary were identified

    3) The 997 most frequent words in the new word list were selected for vocabulary expansion (the original BERT vocabulary permits a maximum of 997 new words).

To compare the effect of using the cancer-domain customized vocabulary on the CancerBERT performance, we pre-trained three CancerBERT models with different sets of vocabulary (described below) based on BlueBERT model:

- CancerBERT$_{OrigVoc}$: used the original BERT vocabulary
- CancerBERT$_{CustVoc\_397}$: used the original BERT vocabulary + 397 cancer-related words based on domain knowledge
- CancerBERT$_{CustVoc\_997}$: used the original BERT vocabulary + 997 cancer-related words based on the term frequency

*Fine-tuning CancerBERT models along with other BERT-based models*

BERT-based models could be further fine-tuned on annotation data to solve specific downstream tasks. The extraction of breast cancer-related phenotypes from texts can be framed as an NER task. The NER is one of the most important tasks in information extraction of text data. It classifies every token in the text into pre-defined entity classes [28]. In this study, we have eight types of name entities: *hormone receptor type, hormone receptor status, tumor size, tumor site, cancer grade, histological type, cancer laterality* and *cancer stage*. The annotated training set was used for the BERT-based model fine-tuning. For major hyperparameters, the max sequence length was set to 128, the training batch size was set to 32 and training epoch was set to 10. The hyperparameters were chosen based on the memory and computing power of our GPU resources. We fine-tuned our CancerBERT$_{OrigVoc}$, CancerBERT$_{CustVoc\_397}$ and CancerBERT$_{CustVoc\_997}$ models, along with the original BERT-large [15], BioBERT [16], BlueBERT [17], CharBERT [22] and character-BERT [23] models on the NER task. All the BERT-based models in this study are uncased. The original BERT-large model was pre-trained on Wikipedia and BookCorpus [15]. The BioBERT model was pre-trained on PubMed abstract and PMC full articles that contain about 18 billion words [16]. BlueBERT model was pre-trained on PubMed abstract and MIMIC-III that contain about 4.5 billion words [17]. The CharBERT was pre-trained on Wikipedia corpus that contains 2.5 billion words [22]. Character-BERT was pre-trained on Wikipedia corpus then further pre-trained on MIMIC-III clinical notes and PMC OA biomedical paper abstracts [23].

**Evaluation**

We evaluated the three CancerBERT models along with other models on the NER task for cancer phenotyping extraction. We applied name entity level evaluation for the NER task. 20% of the annotated data were used as a test set. The F1 score was used as an evaluation metric. For overall performance, the micro-average F1 (calculating precision and recall by counting the sums of the true positives, false negatives, false positives for all classes, and then calculating F1 score) and macro-average F1 (arithmetic mean of all per-class F1 scores) were used. Following the n2c2 evaluation metrics [29], we evaluated in both exact match and lenient match ways for name entities. For exact match, the entity boundary of predicted entity and gold standard should be same, while for lenient match, the entity boundary of predicted entity and gold standard can be overlapped.

We developed BiLSTM-CRF models as the baseline models for the NER task comparison. The input features for the BiLSTM-CRF models are the pre-trained word embeddings. We compared the four different word embeddings (i.e., Word2Vec model pre-trained on Google News [30], Word2Vec model pre-trained on our breast cancer corpus, Global Vectors for Word Representation (GloVe) model pre-trained on Wikipedia [31], and GloVe model pre-trained our breast cancer corpus) as features and finally chose the GloVe trained on Wikipedia corpus [31] as the baseline for further comparison since it obtained the best performance among the four word embeddings for cancer phenotype extraction task.

**RESULTS**

**Performance comparison of CancerBERT models with other BERT models on the cancer phenotyping NER task**

The evaluation results for CancerBERT and other BERT-based models pre-trained in the general biomedical and clinical corpora are shown in **Table 3**. The strict and lenient match F1 scores are shown in the table (lenient match F1 in parenthesis). The scores were averaged scores based on 10 runs, numbers in bold indicate the highest score and asterisk indicates the number is statistically higher than other methods (CI: 0.95). All three CancerBERT models outperformed the baseline models and other BERT models. The CancerBERT$_{CustVoc\_397}$ model obtained the best performance on 4 of 8 entities and obtained the best overall macro F1 scores (0.876 for strict match and 0.904 for lenient match) and micro F1 scores (0.909 for strict match and 0.933 for lenient match). Overall, the performance of CancerBERT$_{CustVoc\_397}$ model significantly surpassed other models; the CancerBERT$_{CustVoc\_997}$ model also obtained good performance.

Table 3– BERT fine-tuning NER entity level evaluation by exact match F1 score (lenient match F1 scores are shown in parenthesis)

| Entity type | BiLSTM-CRF | BERT-large Origin | BlueBERT (PubMed+ MIMIC III) | BioBERT (PubMed) | CharBERT (Wiki) | character-BERT (Medical) | CancerBERT$_{OrigVoc}$ (EHRs corpus) | CancerBERT$_{CustVoc\_997}$ (EHRs corpus) | CancerBERT$_{CustVoc\_397}$ (EHRs corpus) |
|---|---|---|---|---|---|---|---|---|---|
| Hormone Receptor type | 0.953 (0.957) | 0.976 (0.985) | 0.979 (0.984) | 0.982 (0.987) | 0.982 (**0.988**) | 0.972 (0.983) | **0.984** (**0.988**) | 0.979 (0.985) | 0.982 (0.985) |
| Hormone Receptor status | 0.856 (0.856) | 0.846 (0.846) | 0.885 (0.885) | 0.859 (0.859) | 0.878 (0.878) | 0.851 (0.851) | **0.901*** (**0.901***) | 0.887 (0.887) | 0.891 (0.891) |
| Tumor size | 0.664 (0.709) | 0.663 (0.767) | 0.781 (0.819) | **0.785** (0.821) | 0.727 (0.797) | 0.674 (0.684) | 0.765 (0.813) | 0.784 (0.824) | 0.781 (**0.827***) |
| Tumor site | 0.562 (0.771) | 0.696 (0.769) | 0.711 (0797) | 0.749 (0.799) | **0.761*** (0.806) | 0.688 (0.762) | 0.733 (0.792) | 0.715 (0.787) | 0.727 (**0.824***) |
| Tumor grade | 0.910 (0.910) | 0.857 (0.857) | 0.891 (0.891) | 0.886 (0.886) | 0.856 (0.856) | 0.833 (0.833) | 0.891 (0.891) | 0.898 (0.898) | **0.915*** (**0.915***) |
| Tumor laterality | 0.935 (0.935) | 0.926 (0.926) | 0.931 (0.931) | 0.943 (0.943) | 0.948 (0.948) | 0.934 (0.934) | 0.939 (0.939) | 0.947 (0.947) | **0.953*** (**0.953***) |
| Cancer stage | 0.908 (0.908) | 0.804 (0.804) | 0.870 (0.870) | 0.869 (0.869) | **0.909** (**0.909**) | 0.907 (0.907) | 0.870 (0.870) | 0.885 (0.885) | 0.898 (0.898) |
| Histological type | **0.885*** (0.938) | 0.823 (0.918) | 0.843 (0.922) | 0.855 (0.934) | 0.850 (0.927) | 0.861 (**0.943***) | 0.849 (0.922) | 0.862 (0.937) | 0.862 (0.938) |
| Macro average | 0.834 (0.873) | 0.824 (0.859) | 0.862 (0.887) | 0.868 (0.889) | 0.864 (0.888) | 0.840 (0.862) | 0.867 (0.889) | 0.871 (0.896) | **0.876*** (**0.904***) |
| Micro average | 0.876 (0.905) | 0.873 (0.907) | 0.898 (0.921) | 0.904 (0.926) | 0.899 (0.923) | 0.883 (0.906) | 0.903 (0.925) | 0.906 (0.930) | **0.909*** (**0.933***) |

*Note: The scores were averaged scores based on 10 runs, \* indicates statistically higher than other methods (CI: 0.95).*

We explored the total number of unique annotated tokens for each name entity category and how many of them could be identified in the original BERT vocabulary, customized BERT vocabulary based on frequency (for CancerBERT$_{CustVoc\_997}$ model), and customized BERT vocabulary based on domain knowledge (for CancerBERT$_{CustVoc\_397}$ model). The results are shown in **Table 4**.

**Table 4**. Coverage of unique annotated tokens for different BERT vocabularies stated as token count (percentage of total number of unique annotated tokens)

|  | Total number of unique annotated tokens | Exist in original BERT vocabulary | Exist in customized BERT vocabulary based on frequency | Exist in customized BERT vocabulary based on domain knowledge |
|---|---|---|---|---|
| Hormone receptor type | 33 | 14 (42.4%) | 22 (66.7%) | 26 (78.8%) |
| Hormone receptor status | 11 | 4 (36.4%) | 6 (54.5%) | 8 (72.7%) |
| Tumor size | 160 | 62 (38.7%) | 62 (38.7%) | 62 (38.7%) |
| Tumor site | 146 | 88 (60.3%) | 95 (65.1%) | 95 (65.1%) |
| Tumor grade | 20 | 15 (75.0%) | 15 (75.0%) | 18 (90.0%) |
| Tumor laterality | 10 | 4 (40.0%) | 6 (60.0%) | 8 (80.0%) |
| Cancer stage | 58 | 12 (20.7%) | 18 (31.0%) | 52 (89.7%) |
| Histological type | 72 | 28 (38.9%) | 53 (73.6%) | 58 (80.6%) |
| Total | 426 | 178 (41.8%) | 227 (53.3%) | 274 (64.3%) |

We extracted the word embeddings of all unique tokens for each name entity category from the CancerBERT$_{CustVoc\_997}$ and CancerBERT$_{CustVoc\_397}$ models and visualized the word embeddings for these tokens in a two-dimensional plot using t-distributed stochastic neighbor embedding (t-SNE) [32]. Examples of token clusters are shown in **Figure 3**. **Figure 3 (a)** and **Figure 3 (b)** show the clusters of *Cancer stage* (e.g., ptis, n2a, pn1a, t1c), *Hormone receptor status* (e.g., er-positive, er-negative, receptor-positive), and *Tumor laterality* (e.g., left-sided, right-sided, b-left) obtained from CancerBERT$_{CustVoc\_397}$ model. **Figure 3 (c)** and **Figure 3 (d)** show the clusters of *Hormone receptor type* (e.g., estrogen, progesterone, her2), *Hormone receptor status* (e.g., equivocal, amplified, nonamplified) and *Histological type* (e.g., dcis, lcis, lobular, adenocarcinoma) obtained from CancerBERT$_{CustVoc\_997}$ model. Some words in the plots are the newly added words that were not in the original BERT vocabulary, for example, "ptis", "n2a", "pn1a" in *Cancer stage;* "er-positive", "er-negative", "receptor-positive" in *Hormone receptor status*; "left-sided", "right-sided" in *Tumor laterality*; "estrogen", "progesterone, "her2" in *Hormone receptor type, and "*dcis", "lcis" in *Histological type.* The visualization of the entire 426 unique tokens is provided in the supplementary file.

Figure 3. Examples of token clusters in the visualization of word embeddings obtained from CancerBERT$_{CustVoc\_397}$ (a-b) and CancerBERT$_{CustVoc\_997}$ (c-d) models using t-SNE.

**DISCUSSION**

Unstructured EHRs data contains valuable information of patients that can be used for clinical decision support, translational research. In our breast cancer patient corpus, each patient has about 60 pathology reports and over 200 clinical notes. The density of the targeted information is relatively low. As shown in **Table 2**, all name entities have been annotated for more than 100 cases. The *Hormone receptor type, Cancer laterality* and *Histological type* are the most frequent entities with more than 1,000 cases. The language usage for some entities are uniform, only several unique entities exist. For example, for the 1,241 *Cancer laterality* cases, most of them are either "left" or "right". It is relatively easy for the NER models to identify those uniform entities. Some entities are various in the clinical texts. For instance, the *Tumor site* and *Tumor size* have 173 and 305 unique expressions, respectively, in the annotated data, which make them relatively difficult to extract for all models.

For the NER task, the BiLSTM+CRF model outperformed BERT-large origin model for several phenotypes. All other BERT models trained on clinical domain corpus significantly outperformed the BERT-large origin model and baseline BiLSTM-CRF models. In this study, all three CancerBERT models outperformed the baseline model and other state-of-the-art BERT models. It is within expectation that pre-training BERT models on domain-specific corpus could improve the performance of downstream tasks. In addition, we found that adding the cancer domain specific words to the dictionary of the CancerBERT model can further improve the performance. In this study, we applied two methods to add cancer domain-specific words to expand the original BERT vocabulary, adding words based on domain knowledge or frequency. Both methods improve the performance for the cancer phenotype extraction task. Two character-based BERT models (character-BERT and CharBERT) [22-23] were also evaluated and compared with our models for the NER task. The character-based BERT models do not tokenize each word into sub-words; instead, they use additional layers (e.g., CNN, gated recurrent unit) to represent each word using character-level embeddings to avoid the OOV issue. Though the character-based BERT models may improve the robustness (e.g., better handle misspelling issues) compared to the word-based BERT models, they are relatively slower to pre-train and need more computing sources. Our CancerBERT$_{CustVoc\_397}$ and CancerBERT$_{CustVoc\_997}$ models both significantly outperformed the character-based BERT models for the cancer phenotype extraction task, which indicate the advantage of using domain specific vocabulary for specific downstream tasks compared to the character-based BERT models. **Table 4** indicates that both methods to build customized vocabularies could better cover the unique tokens for all name entity categories (except *Tumor size*) compared to the original BERT vocabulary. The domain knowledge-based vocabulary expansion approach covers more tokens compared to the frequency-based expansion approach. **Figure 3** visualizes partial word embeddings of the annotated tokens obtained from the CancerBERT$_{CustVoc\_397}$ and CancerBERT$_{CustVoc\_997}$ models and clear clusters of different cancer name entity categories could be identified from the figure. It indicates that the pre-trained CancerBERT models with customized vocabulary could capture the semantics of different name entities.

We also analyzed the prediction errors produced by our CancerBERT$_{CustVoc\_397}$ model and found that there are mainly three error types. The first is boundary mismatch, which usually happens when the name entity contains multiple tokens, e.g., *Tumor size* and *Tumor site*. For example, in a sentence "The tumor measures 2 cm in length and 1 cm in width", the whole *Tumor size* entity is "2 cm in length and 1 cm in width", but our model only captured "2 cm in length". The second type of error is missing (false negative). For example, "mx" should be predicted as *Cancer stage*, but it was predicted as label "O". And the third error type is mixing up the entities. For example, the number "3" could refer to both *Cancer stage* and *Cancer grade*, sometimes the model could not differentiate them.

This study has certain limitations. We trained the CancerBERT models with customized vocabulary using the breast cancer patient narrative corpus extracted from the EHRs. The corpus contains 5.8 million documents (1 billion tokens); however, it was extracted from a single hospital (UMN), the corpus may not be comprehensive enough to reflect all characteristics of clinical narratives. Another limitation is that all the models were only evaluated on our NER task. In the future, we will further improve our model by integrating corpus from other healthcare institutions and evaluate its generalizability. We will also try different methods (e.g., using MedSpaCy [33] for tokenization) to generate new words. We plan to annotate more data to evaluate our models on other downstream clinical NLP tasks, such as relation extraction and text classification.

## CONCLUSIONS

In this study, a CancerBERT model and its two variations with cancer domain vocabulary were developed to extract the eight breast cancer-related phenotypes from clinical notes and pathology reports in the UMN EHRs. They all outperformed all other existing models; the best model had average macro F1 scores of 0.876 for exact match and 0.904 for the lenient match. We also validated that customized vocabulary may further improve the performance of domain specific BERT models in clinical NLP tasks.

## ACKNOWLEDGEMENTS
We thank Dr. Rubina Rizvi for guiding the annotation process, and thank Ms. Lindsay B. Nichols for proofreading the draft.

## DATA AVAILABILITY
The data underlying this article cannot be shared publicly due to the privacy of patient health information. The pre-trained CancerBERT models will be available at https://github.com/zhang-informatics/CancerBERT after the UMN approval.

## FUNDING STATEMENT
This work is partially supported by the National Center for Complementary and Integrative Health (NCCIH) under grant number R01AT009457 (Zhang); and the University of Minnesota Clinical and Translational Science Institute (CTSI), supported by the National Center for Advancing Translational Sciences under grant number UL1TR002494. The content is solely the responsibility of the authors and does not necessarily represent the official views of the NIH.

## COMPETING INTERESTS STATEMENT

There are no competing interests.

**CONTRIBUTIONSHIP STATEMENT**

SZ conducted the main experiments. NW, SZ and LW participated in the data collection and annotation. HL and RZ guided the study design, data collection and analysis. All authors participated in the writing of the manuscript and critical revisions of the manuscript for important intellectual content.

**Figure captions:**

Figure 1. Examples of annotation in INCEpION.

Figure 2. The training process of CancerBERT models. The CancerBERT models were pre-trained based on the BlueBERT model. The process in the red box was implemented in this study.

Figure 3. Examples of token clusters in the visualization of word embeddings obtained from CancerBERT$_{CustVoc\_397}$ (a-b) and CancerBERT$_{CustVoc\_997}$ (c-d) models using t-SNE.

**Abbreviations:**

HER2: human epidermal growth factor receptor 2

ER: estrogen receptor

EHRs: electronic health record systems

NLP: natural language processing

IAA: inter-annotator agreement

CNN: convolutional neural network

BERT: bidirectional encoder representations from transformers

NER: name entity extraction

i2b2: Informatics for Integrating Biology and the Bedside

UMN: University of Minnesota

CDR: Clinical Data Repository

IRB: Institutional Review Board

CRF: conditional random field

BiLSTM: bidirectional long short-term memory

RNN: recurrent neural networks (RNN)

LSTM: long short-term memory

GloVe: Global Vectors for Word Representation

OOV: out-of-vocabulary